\documentstyle[preprint,aps,floats,psfig]{revtex}
\begin{document}
\newenvironment{tab}[1]
{\begin{tabular}{|#1|}\hline}
{\hline\end{tabular}}

\title {Proximity effect in ferromagnet-
superconductor hybrid structures: role of the
pairing symmetry
}
\author{N. Stefanakis and R. M\'elin}
\address{Centre de Recherches sur les Tr\`es Basses Temp\'eratures,
Centre National de la Recherche Scientifique,
25 Avenue des Martyrs, BP 166, 38042
Grenoble c\'edex 9 France}  
\date{\today}
\maketitle

\begin{abstract}
The spatial variations of the pair amplitude,
and the local density of
states
in $d$-wave or $s$-wave superconductor
-ferromagnet hybrid structures are
calculated self consistently using the Bogoliubov-deGennes formalism within
the two dimensional extended Hubbard model.
We describe the proximity effect in superconductor~/ 
ferromagnet (SF) 
bilayers, FSF trilayers, and interfaces between a superconductor 
and a ferromagnetic domain wall. We investigate in detail the role
played by the pairing symmetry, the  exchange field,
interfacial scattering and crossed Andreev reflection.
\end{abstract}
\pacs{}

\section{Introduction}

The determination of the physics associated to $d$-wave symmetry
has become one of the 
main aspects in the research on high temperature superconductors
\cite{hilgenkamp}.
Tunneling conductance experiments report the existence of a zero bias
conduction peak (ZBCP) \cite{guillou}. 
The origin of the experimental 
ZBCP is explained 
in the context of zero energy states (ZES) formed near 
the $[110]$ surfaces of 
$d$-wave superconductors \cite{stefan,zhu1}. 
These ZES do not appear for $s$-wave 
superconductors or near the $[100]$ surface of $d$-wave superconductors 
and are one of the features that characterize $d$-wave 
superconductors.

In junctions made of $s$-wave or $d$-wave superconductors
connected to normal metals or ferromagnets,
superconducting correlations penetrate in the normal metal 
or in the ferromagnetic region, which is called the proximity effect.
Landauer formalism or quasiclassical models have
been used to calculate 
the tunneling conductance at interfaces between superconductors 
and ferromagnetic metals 
\cite{melin,stefan2,zhu2,kashiwaya}. 
One of the remarkable effects 
is the suppression of the tunneling conductance 
by spin polarization in the ferromagnet \cite{jong}. 
This effect is related to the fact that Andreev reflection
in a multichannel FS point contact takes place only in the
channels having both a spin-up and a spin-down Fermi surface.
In the studies in Refs.~\cite{melin,stefan2,zhu2,kashiwaya}
a simple 
step function variation for the order parameter is assumed, and 
the proximity effect is thus ignored. 

For $s$-wave pairing, 
the proximity effect has been studied in the dirty limit where the 
electron mean free path is shorter than the coherence 
length, by solving 
the Usadel equations \cite{buzdin,baladie}.
It has been demonstrated that in a ferromagnetic metal 
near the boundary with a superconductor the local
density of states (LDOS) at energies close to the Fermi energy $E_F$ 
has a damped-oscillatory behavior similar to
the decay of the Cooper's pair density. 
This effect has been observed experimentally \cite{kontos}.
Moreover transport in SFS trilayers has been
investigated experimentally \cite{bourgeois}.
For cuprate superconductivity, 
spin polarized transport experiments suggest
strong effects due to spin polarization
under the form of a
suppression of the critical current \cite{vasko,fu}.
The extended Hubbard model has already
been used to study the proximity effect
and quasiparticle transport in 
ferromagnet-$d$-wave-superconductor junctions \cite{zhu3}. 
It was found
that the proximity induced $d$-wave pair amplitude oscillates 
in the ferromagnetic region like in the case of
$s$-wave pairing. The tunneling conductance
was also discussed in the framework of a scattering
approach~\cite{zhu3}.

In this paper our goal is to explore several new aspects
related to the proximity effect in 
superconductor ferromagnet (SF) bilayers,
FSF trilayers and interfaces between a superconductor and a
ferromagnetic domain wall (DW).
The spatial variation of the pair amplitude
and the local density of states 
are studied as a function of 
several relevant parameters:
the distance from the surface, the exchange field, the barrier strength, 
and  the symmetry of the 
pair potential. The method is based on exact diagonalizations
of the Bogoliubov-de Gennes equations associated to the mean field
solution 
of an extended Hubbard model.
Our predictions from the simulations of this model are of interest in
view of future STM spectroscopy experiments on hybrid
structures. 

The pair amplitude oscillates 
in the ferromagnetic region both for $s$-wave and $d$-wave,
and the period of oscillations 
decreases with increasing the exchange field. 
The LDOS in the superconductor 
shows large residual values within the gap due to 
the proximity effect, which are increased by the increase 
of the exchange field. In the ferromagnet the proximity effect 
appears as a peak in the LDOS
for $s$-wave and mini-gap for $d$-wave. For both cases the
overall LDOS 
in the ferromagnet is suppressed with the exchange field.

For the $d$-wave superconductor~/ ferromagnet bilayer, 
zero energy states exist only if there exists a finite
barrier potential. This is because zero energy states
are due to back-scattering at the interface in the
presence of a sign change in the order parameter.
The presence of a strong barrier decouples 
the superconductor from the ferromagnet and therefore suppresses
the proximity effect. 
For a S/DW interface
the zero energy peak (ZEP) in the LDOS appears also in
the $d$-wave case 
due to the constructive interference of the 
zero energy states that originate from each domain. 
For FSF trilayers the pair amplitude is larger in the antiferromagnetic 
alignment of the magnetizations of
the ferromagnetic electrodes, both in the
$s$-wave and $d$-wave cases. This can be contrasted with 
recent discussions based on different models~\cite{zirari,baladie}.
We provide
here simple physical arguments to explain 
our numerical results in terms of a) local and non-local Andreev 
reflections and b) pair breaking effects.
 
The article is organized as follows. In Sec. II we
develop the model and discuss the formalism. In Sec. III we discuss
the effect of the exchange field. In Sec. IV we discuss
the effect of the strength of the barrier. 
In Sec. V we discuss the effect of an inhomogeneous exchange field.
In Sec. VI we present the ferromagnet superconductor ferromagnet 
trilayer. Finally
a summary and discussions are presented in the last section.

\section{BdG equations, for the superconductor-ferromagnet junction 
within the Hubbard model} 

The Hamiltonian of
the extended Hubbard model on a two dimensional square
lattice takes the form
\begin{eqnarray}
H & = & -t\sum_{<i,j>\sigma}c_{i\sigma}^{\dagger}c_{j\sigma} 
+\mu \sum_{i\sigma} n_{i\sigma}+\sum_{i\sigma} \mu_i^In_{i\sigma} 
+\sum_{i\sigma} h_{i\sigma}n_{i\sigma} \nonumber \\
  & + & V_0\sum_{i} n_{i\uparrow} n_{i\downarrow}
+\frac{V_1}{2}\sum_{<ij>\sigma\sigma^{'}} n_{i\sigma} n_{j\sigma^{'}}\, 
,~~~\label{bdgH}
\end{eqnarray}
where $i,j$ are sites indices and the angle brackets indicate that the 
hopping is only to nearest neighbors, 
$n_{i\sigma}=c_{i\sigma}^{\dagger}c_{i\sigma}$ is the electron number 
operator at site $i$, $\mu$ is the chemical potential.
$h_{i\sigma}=-h_0\sigma_z$ is the exchange field 
in the ferromagnetic region.
$V_0$, $V_1$  
are on site and nearest-neighbor interaction strength. Negative values 
of $V_0$ and $V_1$ mean attractive interaction and positive values mean 
repulsive interaction. When $V_1<0$ 
the pairing interaction gives rise to $d$-wave 
superconductivity in a restricted parameter range \cite{micnas}. 
To simulate the effect of depletion of the carrier density 
at the surface or impurities the site-dependent 
impurity potential $\mu_i^I$ is set to a sufficiently large  value 
at the surface sites. This prohibits electron tunneling over these
sites. Within the mean field approximation Eq. (\ref{bdgH})
reduces  to 
the Bogoliubov deGennes equations \cite{gennes}:
\begin{equation} 
\left(
\begin{array}{ll}
  \hat{\xi} & \hat{\Delta} \\
  \hat{\Delta}^{\ast} & -\hat{\xi} 
\end{array}
\right)
\left(
\begin{array}{ll}
  u_{n \uparrow}(r_i) \\
  v_{n \downarrow}(r_i) 
\end{array}
\right)
=\epsilon_{n\gamma_1}
\left(
\begin{array}{ll}
  u_{n \uparrow}(r_i) \\
  v_{n \downarrow}(r_i) 
\end{array}
\right)
,~~~\label{bdgbdg1}
\end{equation}

\begin{equation} 
\left(
\begin{array}{ll}
  \hat{\xi} & \hat{\Delta} \\
  \hat{\Delta}^{\ast} & -\hat{\xi} 
\end{array}
\right)
\left(
\begin{array}{ll}
  u_{n \downarrow}(r_i) \\
  v_{n \uparrow}(r_i) 
\end{array}
\right)
=\epsilon_{n\gamma_2}
\left(
\begin{array}{ll}
  u_{n \downarrow}(r_i) \\
  v_{n \uparrow}(r_i) 
\end{array}
\right)
,~~~\label{bdgbdg2}
\end{equation}

such that 
\begin{equation}
\hat{\xi}u_{n\sigma}(r_i)=-t\sum_{\hat{\delta}} 
u_{n\sigma}(r_i+\hat{\delta})+(\mu^I(r_i)+\mu)u_{n\sigma}(r_i)+
h_i\sigma_z u_{n\sigma}(r_i),~~~\label{bdgxi}
\end{equation}

\begin{equation}
\hat{\Delta}u_{n\sigma}(r_i)=\Delta_0(r_i)u_{n\sigma}(r_i)+\sum_{\hat{\delta}} 
\Delta_{\delta}(r_i)u_{n\sigma}(r_i+\hat{\delta}),~~~\label{bdgdelta}
\end{equation}
where the gap functions are defined by

\begin{equation}
\Delta_0(r_i)\equiv 
V_0<c_{\uparrow}(r_i)c_{\downarrow}(r_i)>,~~~\label{bdgdelta0}
\end{equation}

\begin{equation}
\Delta_{\delta}(r_i)\equiv 
V_1<c_{\uparrow}(r_i+\hat{\delta})c_{\downarrow}(r_i)>,~~~\label{bdgdeltadelta}
\end{equation}
and where $\hat{\delta}=\hat{x},-\hat{x},\hat{y},-\hat{y}$. Equations 
(\ref{bdgbdg1},\ref{bdgbdg2}) are subject to the self consistency requirements 

\begin{eqnarray}
\Delta_0(r_i) & = & \frac{V_0(r_i)}{2}F_0(r_i)= \nonumber \\
 &  & \frac{V_0(r_i)}{2}\sum_{n} (
u_{n\uparrow}(r_i)v_{n\downarrow}^{\ast}(r_i)\tanh\left(\frac{\beta 
\epsilon_{n\gamma_1}}{2}\right)+ \nonumber \\ 
 &  & u_{n\downarrow}(r_i)v_{n\uparrow}^{\ast}(r_i)\tanh\left(\frac{\beta
\epsilon_{n\gamma_2}}{2}\right) )
,~~~\label{bdgselfD0}
\end{eqnarray}

\begin{eqnarray}
\Delta_{\delta}(r_i) & = & \frac{V_1(r_i+\hat{\delta})}{2}F_{\delta}(r_i)= \nonumber \\
 &  & \frac{V_1(r_i+\hat{\delta})}{2} \sum_{n} 
(u_{n\uparrow}(r_i)v_{n\downarrow}^{\ast}(r_i+\hat{\delta}) \tanh\left(\frac{\beta \epsilon_{n\gamma_1}}{2}\right) + \nonumber \\
 &  & u_{n\downarrow}(r_i+\hat{\delta})v_{n\uparrow}^{\ast}(r_i) )\tanh\left(\frac{\beta 
\epsilon_{n\gamma_2}}{2}\right)).~~~\label{bdgselfDdelta}
\end{eqnarray}

We start with approximate initial conditions for the gap functions
(\ref{bdgselfD0},\ref{bdgselfDdelta}). After exact diagonalizations of Eqs. 
(\ref{bdgbdg1},\ref{bdgbdg2}) 
we obtain $u_{n\sigma}(r_i)$ and
$v_{n\sigma}(r_i)$ and the
eigenenergies $\epsilon_{n\gamma_1},\epsilon_{n\gamma_2}$.
The quasiparticle amplitudes are inserted in Eqs. 
(\ref{bdgselfD0},\ref{bdgselfDdelta}) and new 
gap functions $\Delta_0(r_i)$ and $\Delta_{\delta}(r_i)$ are evaluated. 
We reinsert these quantities into Eq. (\ref{bdgxi},\ref{bdgdelta})
and we proceed until self-consistency is achieved.
We then compute the $d$-wave and the extended $s$-wave gap 
functions given by the expressions \cite{kallin1}:
\begin{equation}
\Delta_d(r_i)=\frac{1}{4}[\Delta_{\hat{x}}(r_i)+\Delta_{-\hat{x}}(r_i)
-\Delta_{\hat{y}}(r_i)-\Delta_{-\hat{y}}(r_i)],~~~\label{bdgdeltad}
\end{equation}
\begin{equation}
\Delta_s^{ext}(r_i)=\frac{1}{4}[\Delta_{\hat{x}}(r_i)+\Delta_{-\hat{x}}(r_i)
+\Delta_{\hat{y}}(r_i)+\Delta_{-\hat{y}}(r_i)].~~~\label{bdgdeltas}
\end{equation}
The pair amplitude for the $s$-wave case is $F_0(r_i)$.
The pair amplitude for the $d$-wave case is given by the 
expression
\begin{equation}
F_d(r_i)=\frac{1}{4}[F_{\hat{x}}(r_i)+F_{-\hat{x}}(r_i)
-F_{\hat{y}}(r_i)-F_{-\hat{y}}(r_i)].~~~\label{bdgFd}
\end{equation}
The local density of states (LDOS) at the $i$th site is given by
\begin{equation}
\rho_i(E)=-\sum_{n\sigma} 
\left [ |u_{n\sigma}(r_i)|^2 f^{'}(E-\epsilon_n) 
+ |v_{n\sigma}(r_i)|^2 f^{'}(E+\epsilon_n) \right ]
,~~~\label{bdgdos}
\end{equation}
$f^{'}$ is the derivative of the Fermi function,
\begin{equation}
f(\epsilon)=\frac{1}{\exp(\epsilon/k_B T) + 1}
.
\end{equation}

\section{effect of the exchange field}

\begin{figure}
\centerline{
\psfig{figure=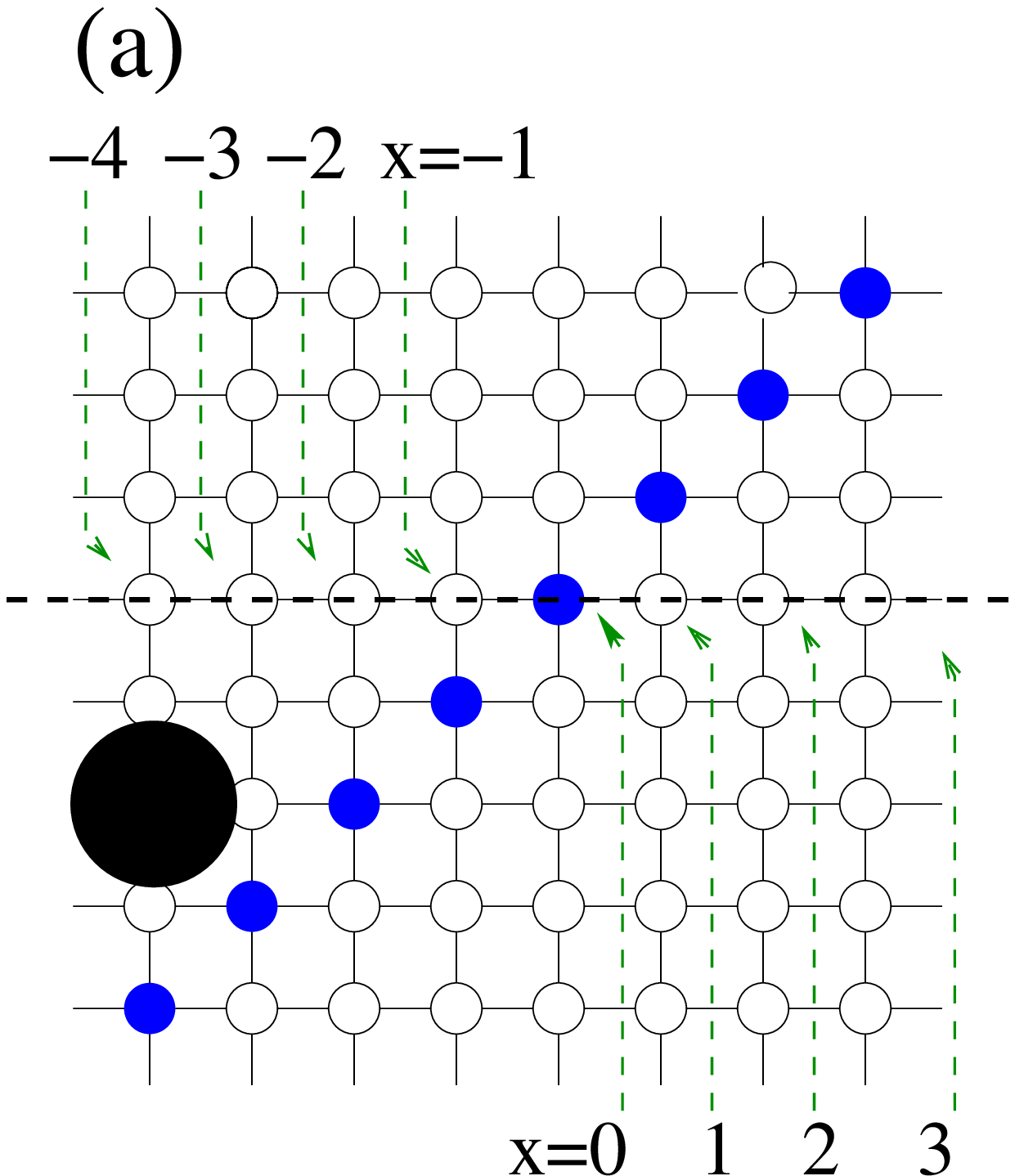,width=5.5cm,angle=0}}
\centerline{
\psfig{figure=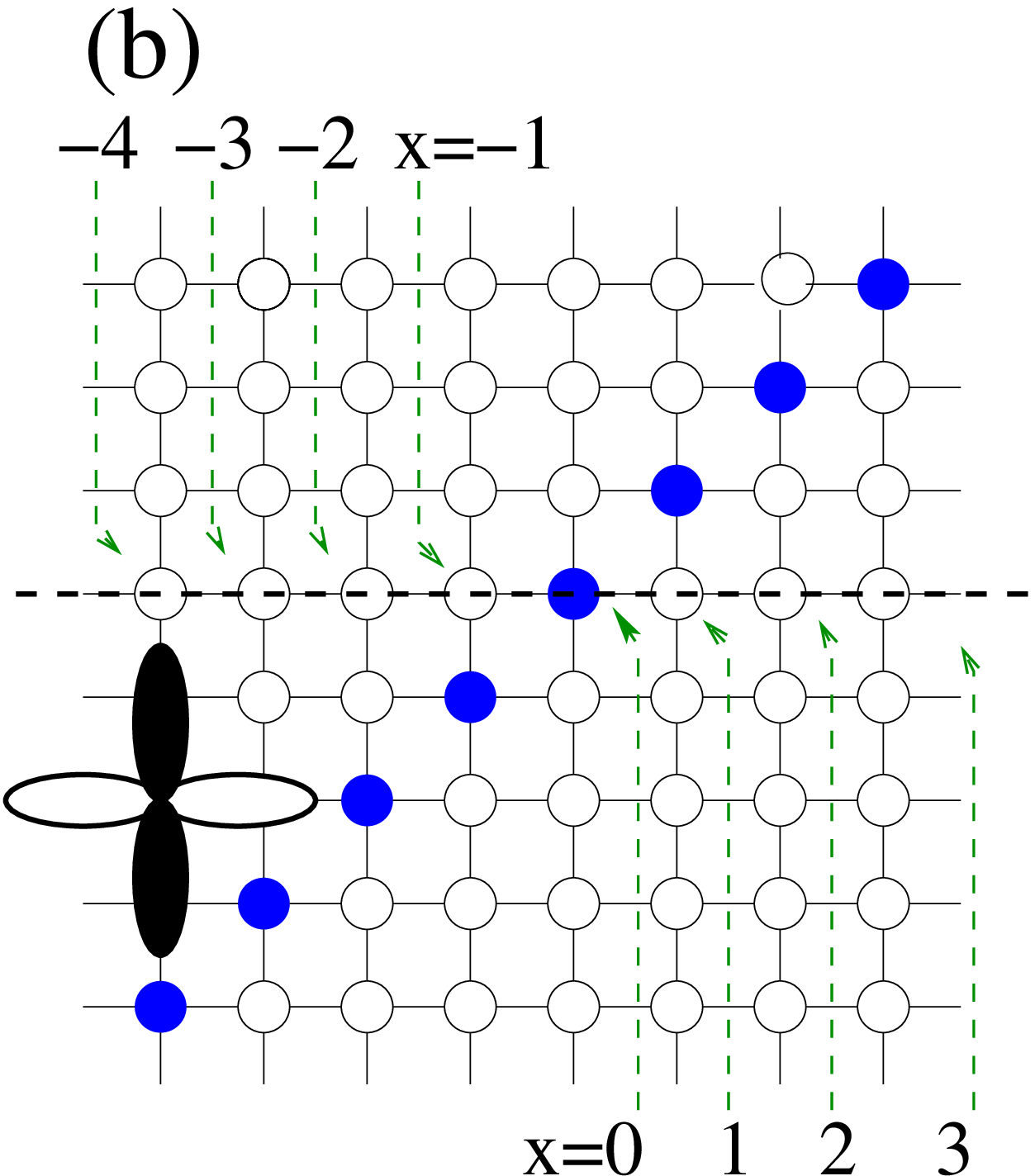,width=5.5cm,angle=0}}
\caption{
(a) Spatial distribution of impurities in the $y'$ direction,
indicated as solid circles, corresponding to
a $s$-wave superconductor-ferromagnet interface along the $[110]$ surface.
The labeling of the sites along the $x$ direction is also shown.
(b) the same as in (a) but for $d$-wave.
The region $x'>0$ represents the ferromagnet and the region
$x'<0$ represents the $s$-wave or $d$-wave superconductor.
}  
\label{is.fig}
\end{figure}

We start our investigation of multiterminal hybrid structures
by discussing the effect of the exchange field on the pair amplitude 
and the local density of states in the cases of $s$-wave
and $d$-wave symmetry.
We consider a two dimensional system of $30\times 30$ sites, and we
impose 
fixed boundary conditions by setting the impurity potential $\mu^{I}=100t$
at the surface.
The temperature is $k_B T=0.1t$.
The interface is modeled by a line of
impurities along the diagonal of the 
lattice, $y'$ direction, where the chemical potential 
on this line of impurities
is set to a value 
related to the strength of the barrier (see Fig.~\ref{is.fig}).
The region $x'>0$ where $V_0=V_1=0$ and $h$ different from zero 
represents the ferromagnet. The region $x'<0$ is considered as the 
$s-$ or $d$-wave superconductor depending on the presence of 
nearest neighbor interaction in the BdG Hamiltonian.

\begin{figure}
\begin{center} 
\leavevmode 
\centerline{\hbox{
\psfig{figure=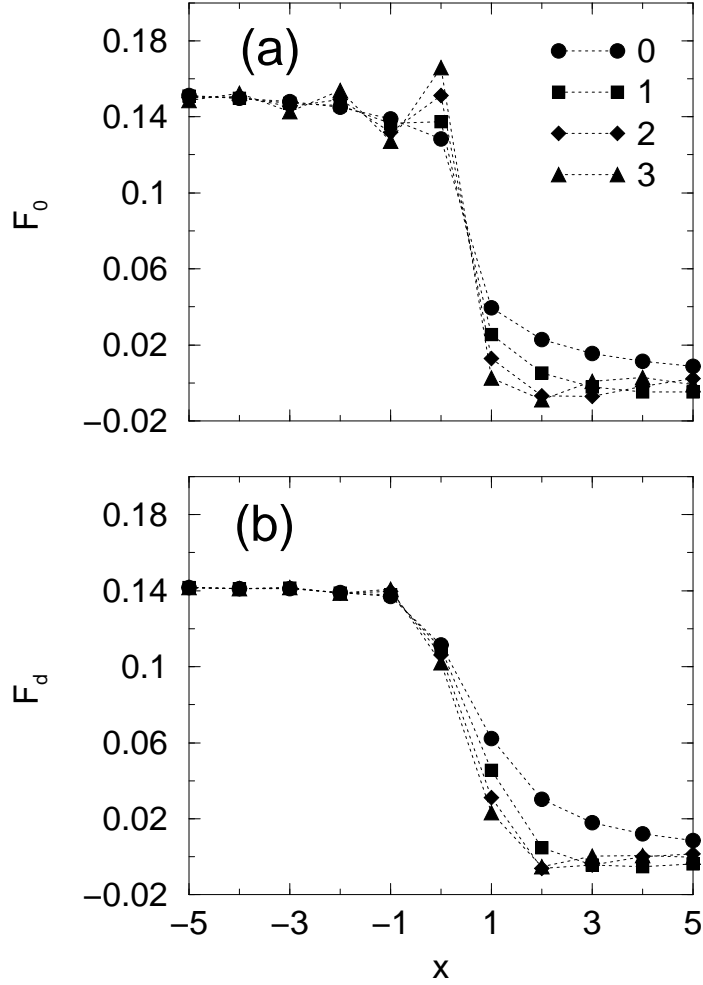,width=8.5cm}}}
\end{center}
\caption{
(a) The $s$-wave
superconducting pair amplitude as a function of $x$, for a $SF$ 
interface for different values of the exchange field $h=0,1,2,3$ (in 
units of $t$).
The pair amplitude is calculated  
along the thick dashed line in direction $x$ shown in Fig. 1. The 
chemical potential in the barrier is $\mu^I=0$.
(b) The same as in (a) but for $d$-wave pairing.
}  
\label{pah.fig}
\end{figure}

Both in $s$-wave and $d$-wave cases, 
the proximity induced pair amplitude for $h=0$
decays monotonically in the normal-metal region and 
oscillates around zero for a finite exchange field 
in the ferromagnetic region.
The period of oscillations decreases 
with increasing the exchange field as seen in Fig. \ref{pah.fig}. 

In the superconducting region the $s$-wave pair amplitude
at the interface is 
enhanced for $h=0$ since the interface is not pair breaking.
As seen in  Fig. \ref{pah.fig}(a) the pair amplitude 
in the superconducting region is modulated by the exchange field
and shows $2 k_F$-oscillations.
In particular the pair amplitude is suppressed by increasing the exchange field
since the interface becomes pair breaking.
Contrary to the $s$-wave case the pair amplitude for the 
$d$-wave case is suppressed for all values of the exchange field 
since the interface itself is pair breaking. Also as visible 
in figure \ref{pah.fig}(b) the pair amplitude is not modulated 
in the superconducting region which should be contrasted with
the $s$-wave case on Fig.~\ref{pah.fig}(a).

\begin{figure}
\begin{center} 
\leavevmode 
\centerline{\hbox{
\psfig{figure=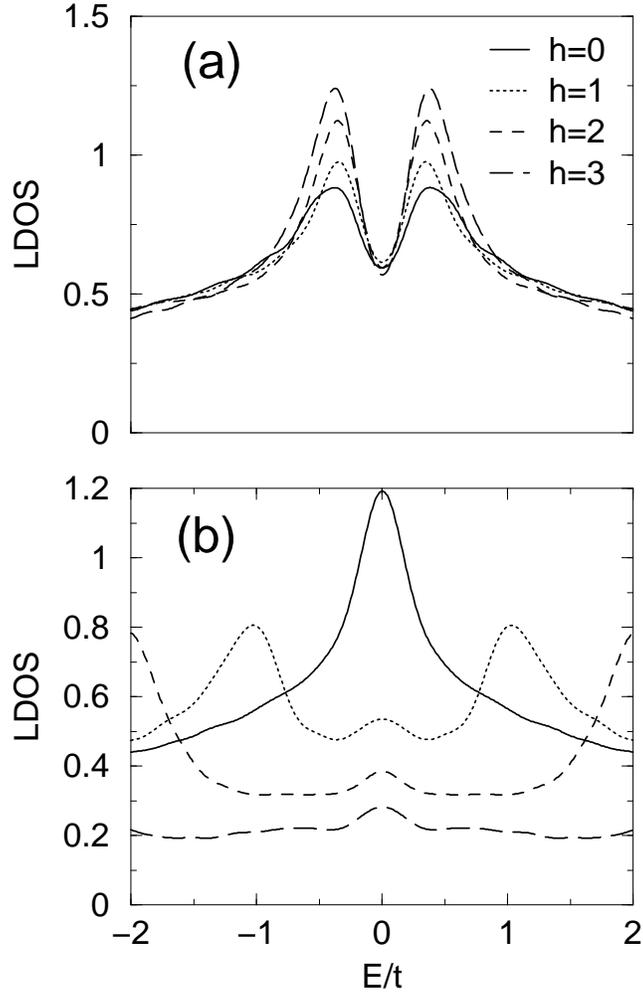,width=8.5cm}}}
\end{center}
\caption{
The local density of states for the $s$-wave case,
as a function of $E/t$ for the site $x=0$ 
in the barrier (a), and $x=1$ in the ferromagnet (b)
shown in Fig. 1, for different values of the exchange field 
$h=0,1,2,3$. The chemical potential in the barrier $\mu^I$ is zero.
}  
\label{ldoss.fig}
\end{figure}

\begin{figure}
\begin{center} 
\leavevmode 
\centerline{\hbox{
\psfig{figure=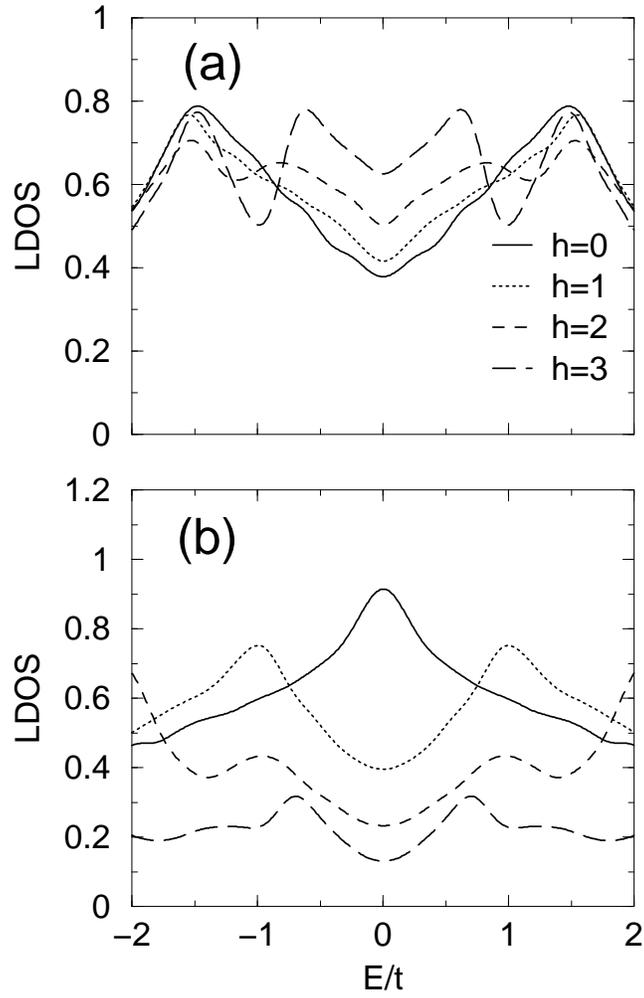,width=8.5cm}}}
\end{center}
\caption{
The same as in Fig. \ref{ldoss.fig} but for $d$-wave pairing state.
}  
\label{ldosd.fig}
\end{figure}

In the $s$-wave case 
the LDOS in the ferromagnet decreases with increasing the 
exchange field as seen in Fig. \ref{ldoss.fig}(b). For 
$h=0$ a single peak exists at $E=0$. 
This peak is due to the shape of the LDOS of the two-dimensional
lattice model that we use in the simulations. There is also
an extra contribution to this peak due to the proximity
effect.
For a finite exchange 
field two peaks exist symmetrically around zero at $E=\pm h$, 
since the spin-up and spin-down bands are shifted with respect to the
Fermi energy in the presence
of the exchange field. These peaks have been observed in STM 
experiments for the proximity effect of magnetic particles of Cobalt
included in a Niobium layer \cite{cretinon}.
The asymmetry observed in the experiment is due to the depletion 
of the chemical potential in the magnetic region. 
The long ranged proximity induced LDOS in the
ferromagnet at the exchange energy has also been found in the 
diffusive limit described by Usadel equations \cite{buzdin}.
Also we find that the zero energy LDOS is enhanced in the
ferromagnet (which we call a ``reversal of the gap region''),
as opposed to the fact that the zero energy
LDOS is reduced in the superconductor. 
The reversal of the gaped region
in the $s$-wave case is in agreement with 
the experiments in Ref.~\cite{kontos}.
In the superconducting region 
the LDOS develops a gap (see Fig. \ref{ldoss.fig}(a)).
However due to the proximity effect 
residual values of the LDOS exist within the gap which 
correspond to the two peaks in the LDOS on Fig.~\ref{ldoss.fig}(a).
The residual 
values of the LDOS depend on the exchange field and have
a larger contribution for a larger exchange field as expected. 
This can be understood from the fact that pair-breaking effects
increase with the exchange field in the ferromagnet.
As a consequence
there is an increase of quasiparticle states within the gap
if we increase the exchange field in the ferromagnet which
constitutes a qualitative explanation to Fig.~\ref{ldoss.fig}(a).
We can provide another qualitative explanation in terms of 
local Andreev reflections in the ferromagnetic electrode 
which decrease as we increase the exchange field. As a 
consequence the transfer of Cooper pairs in the superconductor 
decreases and the LDOS in the superconductor increases.
Both arguments apply also in the $d$-wave case
(see Fig.~\ref{ldosd.fig}).

The presence of residual values in the LDOS due to the proximity effect
is also a property 
of $d$-wave superconductor/ferromagnet interfaces 
as seen in Fig. \ref{ldosd.fig}. 
The line-shape of the LDOS is V in the case of $d$-wave symmetry
whereas the line-shape was U in the $s$-wave case. 
Increasing the exchange field increases 
the residual values in the gap in the superconductor 
(see Fig. \ref{ldosd.fig}(a)).
We see from Fig.~\ref{ldosd.fig}(a) that two energy scales
are involved in the LDOS for $d$-wave symmetry of the
order parameter. For instance on Fig.~\ref{ldosd.fig}(a)
a local maximum of the LDOS is obtained
at high energy for $E/t\simeq 1.5$ and a second
local maximum is obtained at a lower energy for
$E/t\simeq 0.6$.
We see on Fig.~\ref{ldosd.fig}(a) that the low
energy and high energy LDOS react differently to the
increase of the exchange field. Compared to the
$s$-wave case discussed previously there is an
additional pair breaking mechanism in the case
of $d$-wave which is due to the existence of
quasiparticle states within the gap for some
orientations. In the case of the [110] interface
represented on Fig.~\ref{is.fig} and used on
Fig.~\ref{ldosd.fig}(a) it is expected that
spin polarized electrons from the ferromagnet
can penetrate in the superconductor along
the direction for which the gap cancels which
is a pair breaking mechanism not present
for $s$-wave.

On the ferromagnetic side of the interface 
(see Fig.~\ref{ldosd.fig}(b))
we do not observe a reversal of the gaped 
region unlike the case of $s$-wave superconductor
on Fig.~\ref{ldoss.fig}(b).

\begin{figure}
\begin{center} 
\leavevmode 
\centerline{\hbox{
\psfig{figure=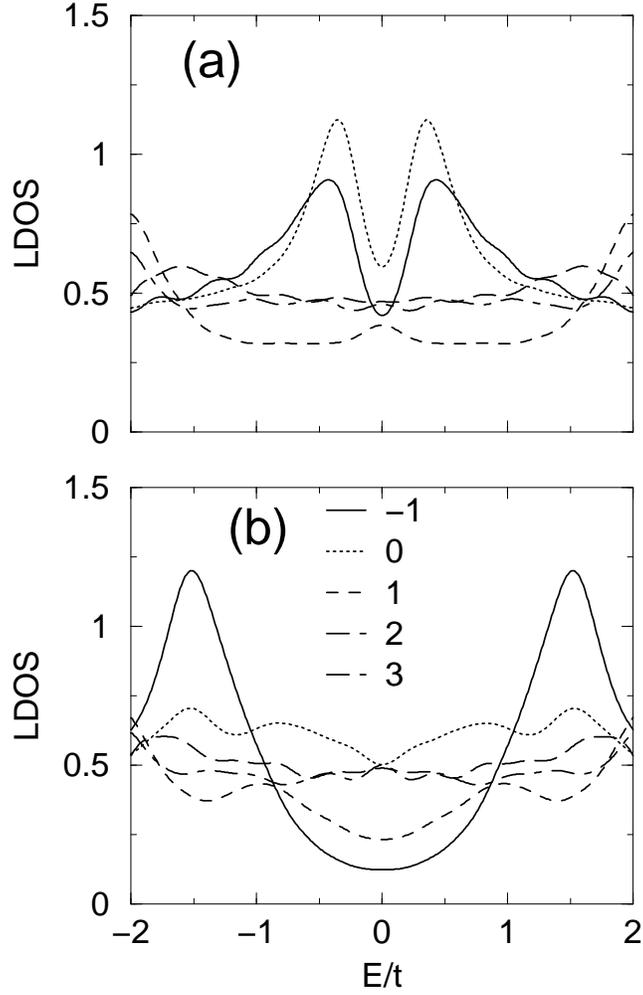,width=8.5cm}}}
\end{center}
\caption{
(a) The local density of states
as a function of $E/t$ for exchange field $h=2$,
for different values of the distance from the interface
$x=-1,0,1,2,3$. The pairing state is $s$-wave.
(b) The pairing state is $d$-wave.
$x=-1$ is a site in the superconductor.
$x=0$ is a site in the barrier.
$x=1,2,3$ are sites in the ferromagnet.
}  
\label{ldosdist.fig}
\end{figure}

The LDOS shows strong modulations with the distance from the 
interface both for $s$ and $d$-wave pairing 
as seen in Fig. \ref{ldosdist.fig}. Increasing 
the exchange field $h$ increases
the residual values of the LDOS in the gap in the 
superconductor. 
In both cases the zero energy 
density of states oscillates as a function of the distance to
interface as found in experiments \cite{kontos}. 
The $d$-wave gap is more sensitive to the proximity effect
of the ferromagnet.

\section{effect of the interfacial scattering potential}
\begin{figure}
\leavevmode 
\centerline{\hbox{
\psfig{figure=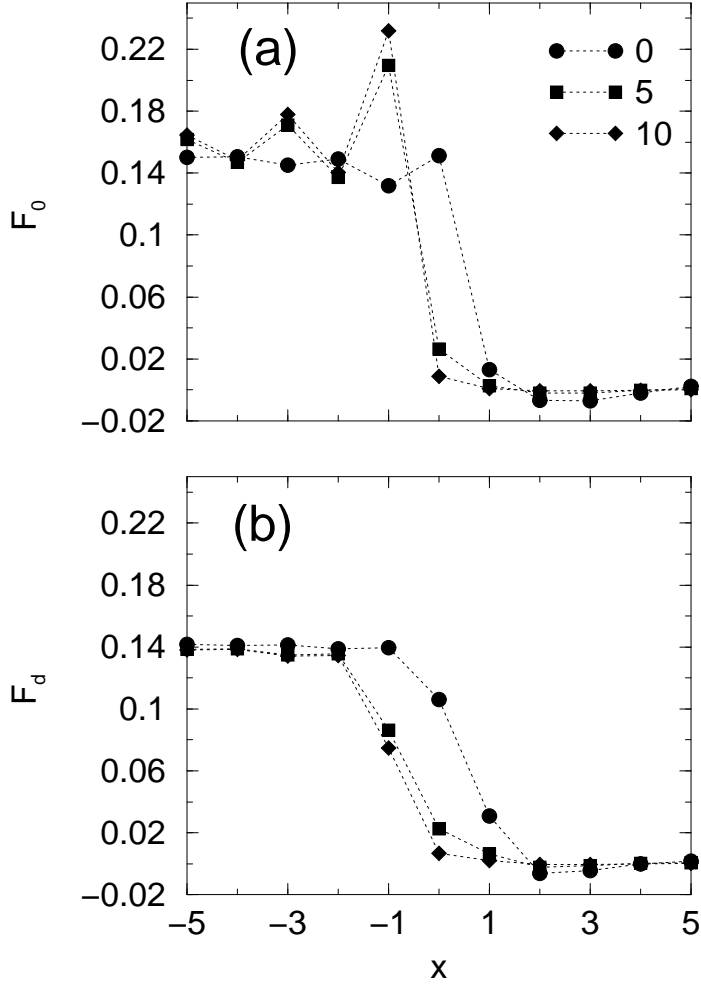,width=8.5cm}}}
\caption{
(a) The $s$-wave
superconducting pair amplitude as a function of $x$, for a $SF$
interface for different values of the interface barrier $\mu^I=0,5,10$ (in
units of $t$).
The pair amplitude is calculated
along the thick dashed line in direction $x$ shown in Fig. 1. The
exchange field in the ferromagnet is $h=2$.
(b) The same as in (a) but for $d$-wave pairing.
}  
\label{pab.fig}
\end{figure}

In general the increase of the interfacial barrier potential
suppresses the 
proximity effect because the leakage of Cooper pairs from the
superconductor to the N or F electrodes is reduced if the
tunnel amplitude is reduced.
As a consequence
for both $s$-wave and 
$d$-wave the oscillations of the pair amplitude are 
reduced on the ferromagnetic side by the increase of the 
barrier strength as seen in Fig. \ref{pab.fig}. 
In the $s$-wave case (see Fig. \ref{pab.fig}(a))
the pair amplitude on the 
superconducting side increases by the increase of the 
barrier strength since the barrier decouples the superconductor 
from the ferromagnet and makes the interface less pair breaking. 
In the $d$-wave case on the 
other hand the pair amplitude close to the interface is 
determined by the competition of two effects:
(i) the decoupling of the 
two layers which makes the interface less pair breaking,
and (ii) the reflection of quasiparticles at
the barrier which makes the interface more pair breaking. 
As seen in Fig. \ref{pab.fig}(b) the latter process 
dominates over the former and the pair amplitude 
decreases with the exchange field on the superconducting
side of the interface.

\begin{figure}
\leavevmode 
\centerline{\hbox{
\psfig{figure=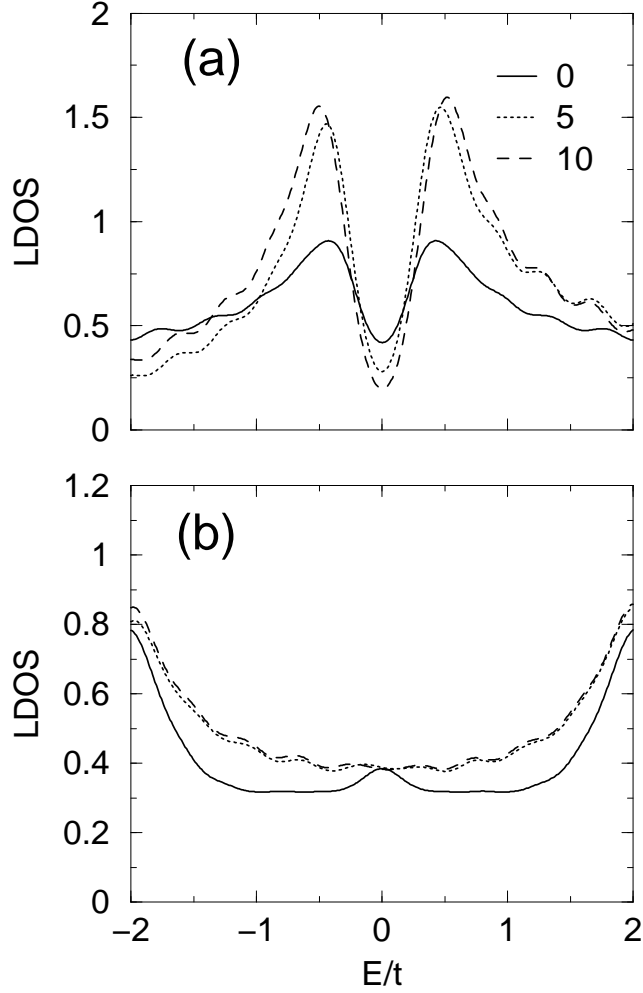,width=8.5cm}}}
\caption{
The local density of states for the $s$-wave case,
as a function of $E/t$ for the site $x=-1$ (a) on the
superconducting side of the interface, and $x=1$ (b)
on the ferromagnetic side of the interface, for different
values of the barrier strength $\mu^I=0,5,10$, $h=2$.
The geometry is represented on Fig. 1.
}  
\label{ldosbars.fig}
\end{figure}

For the $s$-wave case, the subgap LDOS is suppressed by an
increase of the barrier strength and a gap develops on the 
superconducting side (see Fig. \ref{ldosbars.fig}(a)) because
the superconductor decouples from the ferromagnet when
the barrier strength increases. Moreover the Andreev reflection 
process is suppressed.
On the other hand it is visible on Fig.~\ref{ldosbars.fig}(b)
that
the LDOS on the ferromagnetic site 
becomes flat (i.e. the peak at zero energy disappears).
This proves that the ZEP is due to the proximity effect
because the two sides of the interface become decoupled
when the strength of the barrier increases.
The LDOS in the barrier (not presented) 
is suppressed since 
the quasiparticles are not able to tunnel to these sites 
for very large values of the impurity potential.

\begin{figure}
\leavevmode 
\centerline{\hbox{
\psfig{figure=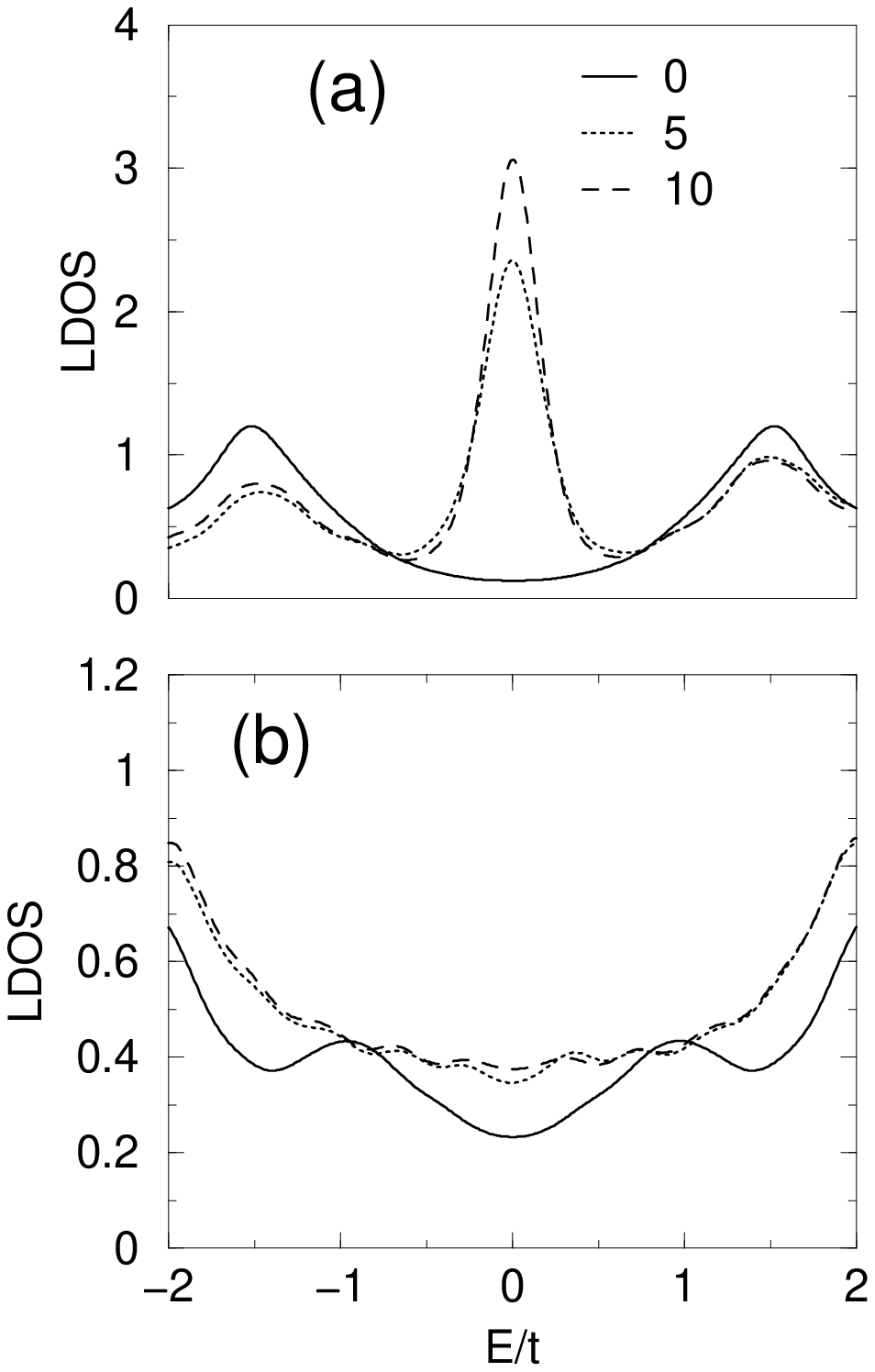,width=8.5cm}}}
\caption{
The same as in Fig. \ref{ldosbars.fig} but for the $d$-wave case.
}  
\label{ldosbard.fig}
\end{figure}

In the $d$-wave case a ZEP develops and 
its height increases with increasing the barrier strength.
Increasing the barrier strength favors the
reflection of quasiparticles at the interface and 
the formation of Andreev bound states due to the sign change 
of the pair amplitude. This is demonstrated in Fig. \ref{ldosbard.fig}(a).
The ZES does not exist when the 
barrier is absent because the quasiparticles are not reflected 
at the interface.  
On the ferromagnetic side the mini gap due to the proximity effect
disappears for large barrier strength (see Fig. \ref{ldosbard.fig}(b)). 

\section{effect of a magnetic domain wall}
\begin{figure}
\centerline{
\psfig{figure=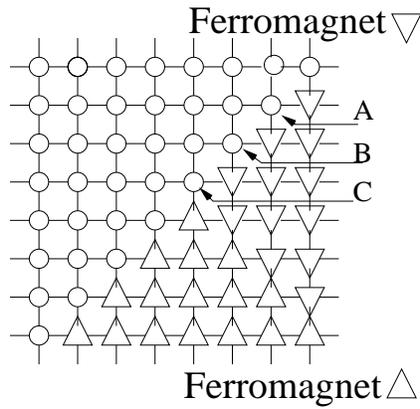,width=5.5cm,angle=0}}
\caption{
Interface between a $s$- or $d$-wave
superconductor and a domain wall
between a ferromagnet with spin-up orientation indicated as up-triangle
and a ferromagnet with spin-down orientation indicated as down-triangle.
The labeling of several sites along the interface is also shown.
}
\label{diagonald.fig}
\end{figure}

We want to study the quasiparticle properties in the presence of 
a magnetic domain wall (DW). 
We suppose that
the DW is formed in a tricrystal geometry where a superconductor 
is in contact with a spin-up and a spin-down ferromagnetic
domain as seen 
in Fig. \ref{diagonald.fig}.
We use here a schematic model in which there is no rotation
of the magnetization inside the DW. Namely the DW
on Fig.~\ref{diagonald.fig} is just a juxtaposition of a
spin-up and a spin-down magnetic domain. It is expected that
this model is a valid description of the regime where the
width of the DW is small compared to the superconducting
coherence length. This hypothesis may not be well
verified for $d$-wave superconductors since the 
superconducting coherence length is very short in
high-$T_c$ materials. Nevertheless even though our model
is not fully realistic for $d$-wave pairing we show that
it contains an interesting physics that can be a useful
guideline for understanding more realistic models in the future.
More precisely we
want to address the question of the 
contribution of crossed Andreev reflection to 
the LDOS and to provide a connection with recent 
experimental data where a similar geometry is studied
with a conventional superconductor~\cite{giroud}.

\begin{figure} 
\leavevmode  
\centerline{\hbox{ 
\psfig{figure=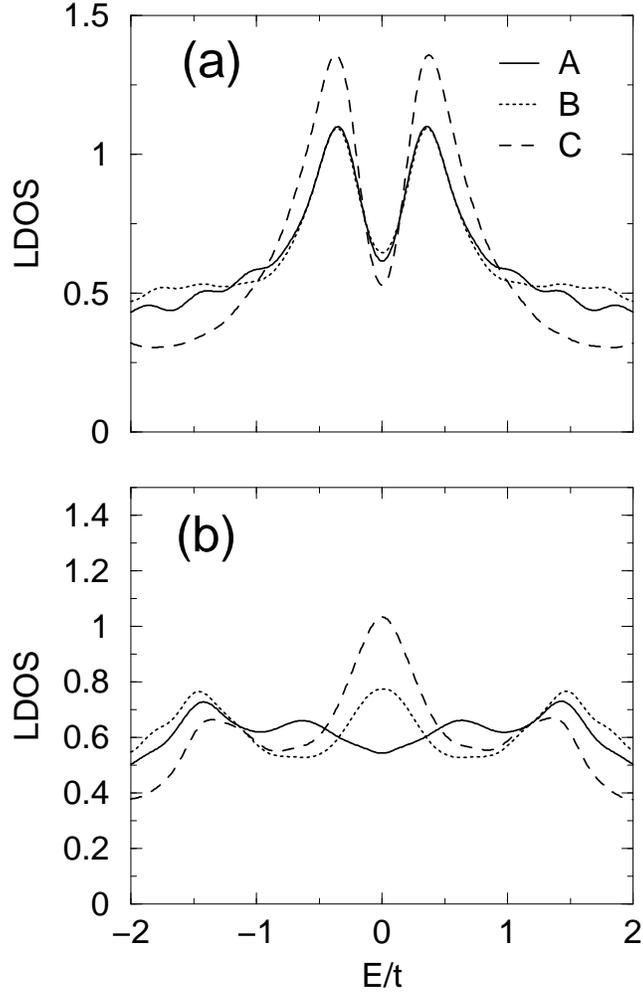,width=8.5cm}}} 
\caption{ 
(a) The local density of states for the $s$-wave case, 
as a function of $E/t$ for the site $A,B,C$ shown in the 
previous figure along the interface. There is no impurity potential 
at the superconductor-ferromagnet interface.
(b) The same as in (a) but for $d$-wave. The ZEP appears only at 
site $C$ for the $d$-wave case.
}   
\label{domain.fig} 
\end{figure}

Due to the presence of crossed Andreev 
reflection we find that in the $s$-wave case the subgap LDOS
is smaller than the
LDOS at the superconductor ferromagnet interface with a single spin 
orientation of the magnetic moment. As we see in \ref{domain.fig}(a) the 
subgap LDOS is smaller at the site in front of the DM due to  
crossed Andreev reflection where an spin-up electron
from one domain 
is reflected as a spin-down hole in the other domain. This 
process transfers a Cooper pair inside the superconductor, and 
enhances superconductivity. As a consequence the 
subgap LDOS is reduced. An equivalent 
explanation can be given in terms of pair breaking close to the DW:
the pairs that are formed in front of the DW are more stable
because they experience the spin-up and spin-down
orientation of the DW. 
Therefore the effects due to pair breaking are small in DW 
compared to SF structure and the subgap LDOS is thus reduced.
This behavior is in agreement with recent study of multiterminal 
hybrid structures \cite{zirari} 
and in agreement with experimental data \cite{giroud}.

In the case of a $d$-wave order parameter
we showed in the previous section that a ZEP is formed at the 
superconductor ferromagnet interface only in the presence of a finite 
barrier strength and $d$-wave symmetry. 
The height of the ZEP increases as a function of the barrier strength.
In the present section we demonstrate 
that also a magnetic inhomogeneity, e.g., a domain wall (DW) 
is sufficient for the 
formation of a ZEP in the LDOS in the absence of impurity potential. 
This is illustrated in Fig. \ref{domain.fig}(b) where the LDOS is 
plotted as a function of energy along the interface sites, for the 
$d$-wave case. 
We see that far away from the DW the LDOS does not show a ZEP. However 
as we approach the DW a ZEP develops.
A first qualitative explanation is crossed Andreev reflection
that is affected by the sign change of the order parameter
which enhances the zero-energy LDOS. This is qualitatively
consistent with Ref.~\cite{stefan4}.
These results can also be well explained in terms 
of quasiparticle subgap states that originate from the local 
suppression of the $d$-wave pairing amplitude very close to
the inhomogeneity. 
In the $s$-wave case which has been discussed in the
previous paragraph, the magnetic inhomogeneity 
increases superconducting correlations. This is modified
in the presence of $d$-wave superconductivity due to the
strong suppression of the order parameter and the formation
of quasiparticle states inside the gap.

\section{FSF trilayer}

\begin{figure} 
\leavevmode  
\centerline{\hbox{ 
\psfig{figure=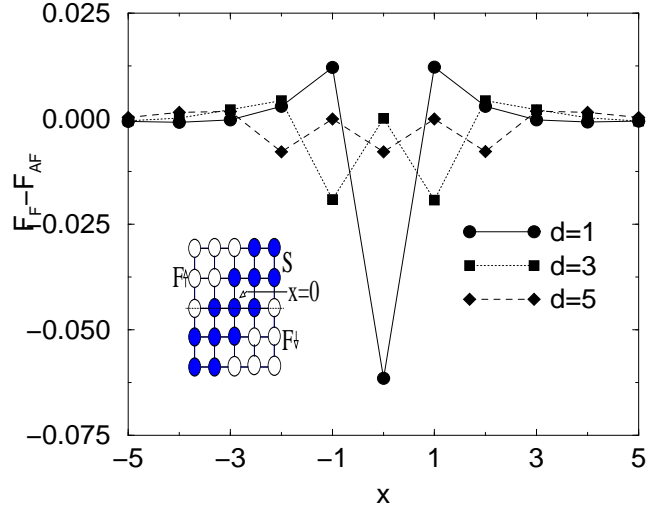,width=8.5cm}}} 
\caption{ 
The difference between the pair amplitude in the 
ferromagnetic and antiferromagnetic 
alignment of magnetizations for the $s$-wave case, 
as a function of the distance from the 
center of the FSF trilayer (of thickness d=3) shown in the 
inset.
}   
\label{newfsf.fig} 
\end{figure}

Now we describe trilayers consisting of 
a superconductor inserted in between two ferromagnets. 
We show that the LDOS and the pair amplitude are controlled by the 
orientation of magnetizations in the ferromagnets. In particular 
we find that
inside the superconductor the pair amplitude is larger in the
antiparallel alignment (see Fig. \ref{newfsf.fig}). 
This result can be contrasted with a recent work~\cite{zirari}
in which is was shown from the solution of an
analytical model
that the superconducting order parameter
is larger in the ferromagnetic alignment.
The self consistent 
approach that we describe here agrees with several analytical
solutions that exist in the literature like the Usadel or the 
Eilenberger equations \cite{buzdin}.

\begin{figure} 
\leavevmode  
\centerline{\hbox{ 
\psfig{figure=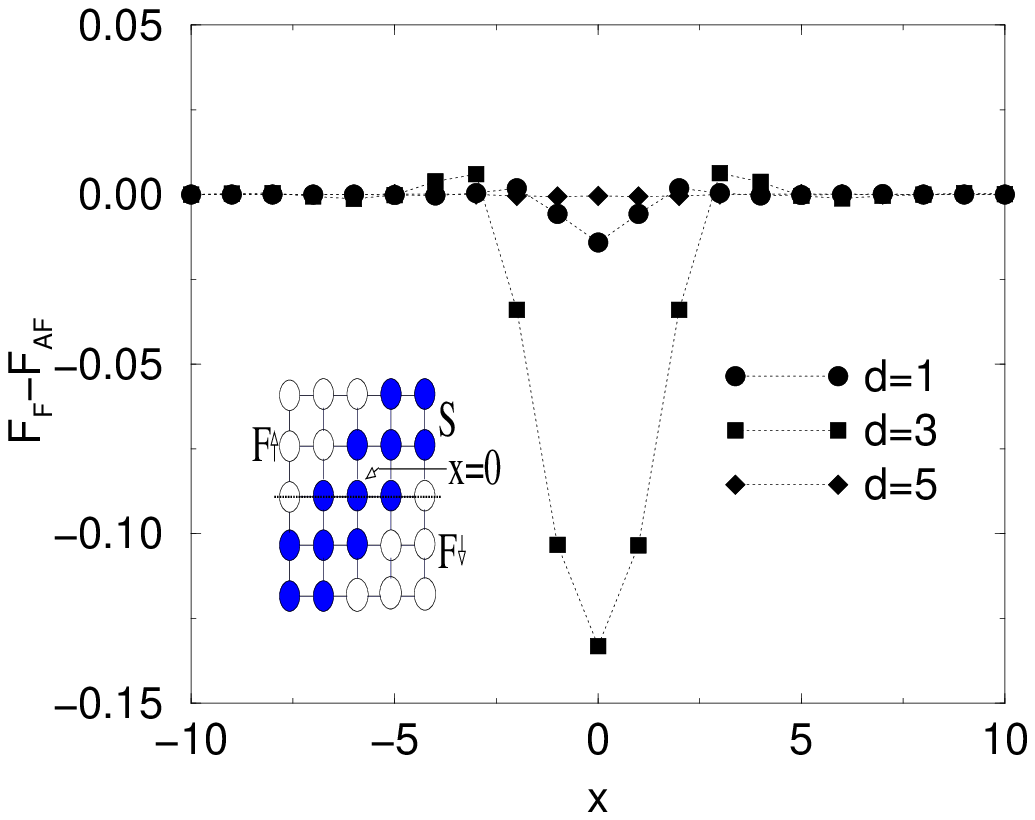,width=8.5cm}}} 
\caption{ 
The same as in figure \ref{newfsf.fig} but for the $d$-wave
case.
}   
\label{newfdf.fig} 
\end{figure}

For the $d$-wave case the interface $(110)$ is additionally 
pair-breaking due to the sign change of the order parameter. 
As a consequence
the effect is much more pronounced than for $s$-wave,
as seen in Fig. \ref{newfdf.fig}. 
Also due to the small coherence length of the $d$-wave superconductor
the effect vanishes as soon as the width $d$ of the superconductor
is larger than $5$ atomic layers in our simulation. 
Note that for $d=1$ we have a very 
special form of $d$-wave superconductivity: since the $d$-wave pairing
involves neighboring sites, normally it is not expected
to appear for a one dimensional 
superconducting chain. However in our case the finite $d$-wave 
order parameter is due to the proximity effect only. 

\begin{figure} 
\leavevmode  
\centerline{\hbox{ 
\psfig{figure=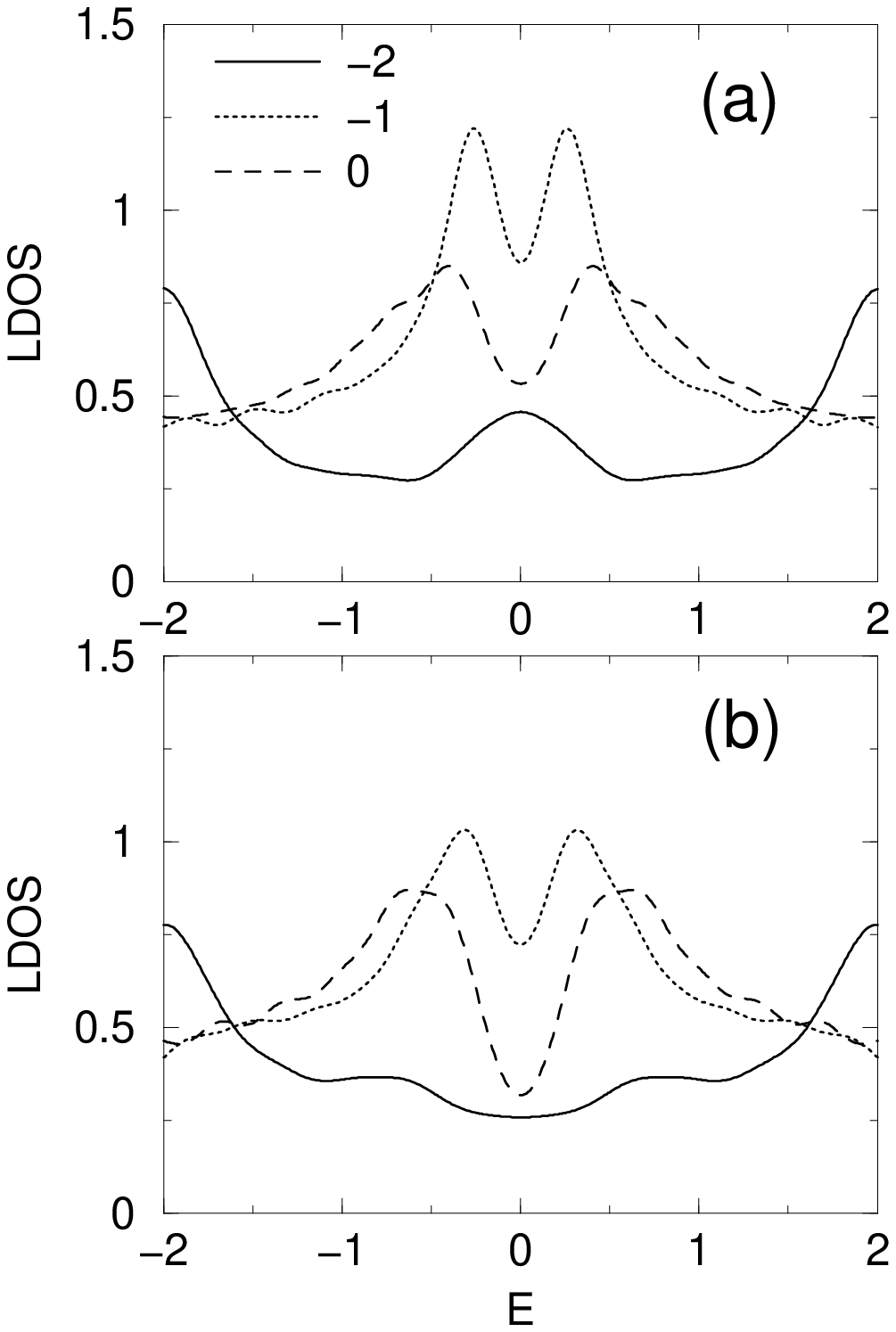,width=8.5cm}}} 
\caption{ 
(a) The local density of states for the $s$-wave case, 
as a function of $E/t$ for the sites at distance $-2,1,0$ from the 
center of the trilayer FSF (of thickness d=3) shown in the 
previous figure along the interface, for the antiferomagnetic 
alignment of magnetization in the ferromagnetic electrodes.
(b) The same as in (a) but for the ferromagnetic alignment.
}   
\label{fsfdos.fig} 
\end{figure}

\begin{figure} 
\leavevmode  
\centerline{\hbox{ 
\psfig{figure=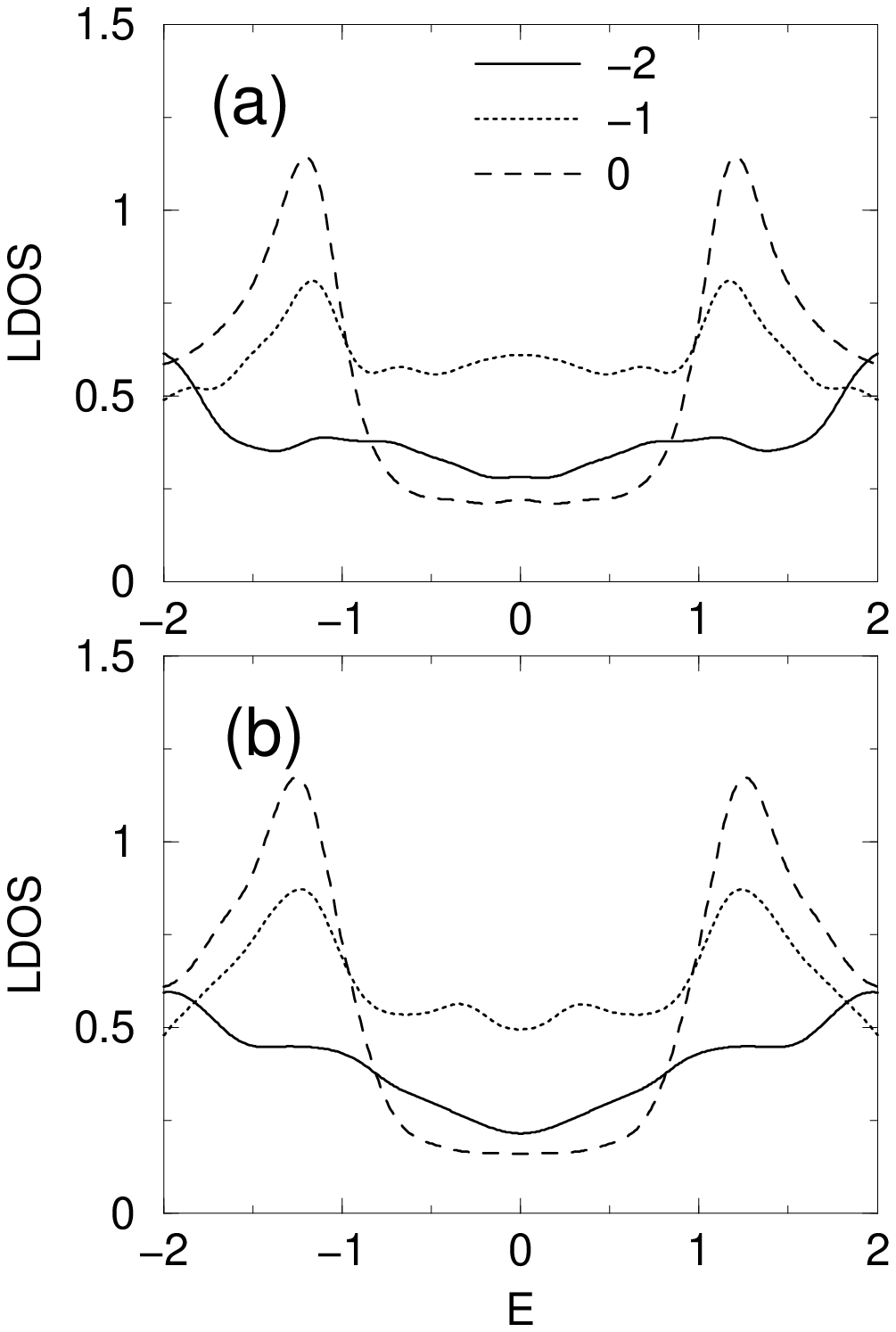,width=8.5cm}}} 
\caption{ 
(a) The local density of states for the $d$-wave case, 
as a function of $E/t$ for the sites at distance $-2,1,0$ from the 
center of the trilayer FSF (of thickness d=3) shown in the 
previous figure along the interface, for the antiferomagnetic 
alignment of magnetization in the ferromagnetic electrodes.
(b) The same as in (a) but for the ferromagnetic alignment.
}   
\label{fdfdos.fig} 
\end{figure}

The LDOS gives an information that can be directly compared
to tunneling conductance measurements 
and STM experiments. For the antiparallel alignment of the spins in the two 
ferromagnetic electrodes the LDOS shows a gaped structure in the 
superconductor which is enhanced near the boundaries due to the 
proximity effect as seen in Fig. \ref{fsfdos.fig}(a). 
In the top sites inside the ferromagnet we see an 
reversal of the gaped structure as in the proximity effect in bilayers.
The same qualitative conclusions hold for the parallel alignment 
but in this case there is no inverse of the gaped region 
in the ferromagnet (see in Fig. \ref{fsfdos.fig}(b)).
Similar effects are observed for the $d$ wave case. Also in this case no
reversal of the gaped region is observed in the ferromagnet 
(see in Fig. \ref{fdfdos.fig}(b)). 

\section{conclusions}
We calculated the LDOS and the pair amplitude for several 
ferromagnet~/ superconductor hybrid structures,
within the extended Hubbard 
model, self consistently. 
In SF bilayers we found
that the proximity induced pair amplitude 
oscillates in the ferromagnetic region. The period of
oscillations decreases with increasing the exchange field.
The pair amplitude is suppressed for the $d$-wave interface 
but is enhanced for the $s$-wave interface. The proximity effect in the 
LDOS is demonstrated as a peak for $s$-wave and a mini-gap for 
$d$-wave.
The LDOS in the superconductor shows large residual
values due to the proximity 
effect which are increased by an increase of the exchange 
field due to the suppression of Andreev reflection. 
The barrier decouples the superconductor from the 
ferromagnet and suppresses the proximity effect. 
The zero energy peak develops in the LDOS only for $d$-wave pairing, 
finite barrier strength and its height increases with the increase 
of the barrier strength. 

The zero energy peak in the LDOS appears also for
the $d$-wave case at the interface of a superconductor with a 
magnetic domain wall due to the constructive interference of the 
zero energy states that originate from each domain.
For the $s$-wave case the subgap LDOS is reduced due to the crossed 
Andreev reflections which transfer Cooper pair in the superconductor.

In FSF trilayers the pair amplitude is larger in the 
antiferromagnetic alignment for $s$-wave. The same is true for 
$d$-wave in a higher scale because of additional pair breaking 
mechanism due to the pairing symmetry.
\bibliographystyle{prsty}

\end{document}